# *Direct Dependence of Covalent, Van Der Waals and Valence Shell Radii of Atoms on Their Bohr Radii for Elements of Groups 1A – 8A

Raji Heyrovska

Institute of Biophysics of the Academy of Sciences of the Czech Republic,

Královopolská 135, 612 65 BRNO, Czech Republic.

Email: rheyrovs@hotmail.com

Abstract

Recent work by the author has shown that ionic, atomic and the ground state Bohr radii ($a_B$) of elements are inter-related. An earlier work by others has shown that the ratio, van der Waals radii/de Broglie wavelengths is nearly constant for each group of some non-metallic elements. Since the Bohr radius and the de Broglie wave length are directly related, the author undertook to investigate the results obtained by the earlier researchers. This work shows (for the first time) that in fact for all the elements of groups 1A – 8A, 1) the van der Waals radii are directly proportional to the Bohr radii, 2) the valence shell and covalent radii also vary linearly with the Bohr radii and that 3) all the above radii (R) can be unified by a single linear equation, $R = ma_B + c$. Therefore, the Bohr radius can be considered as a unit of length for the above radii as much as for the smaller Compton wavelength ($\alpha a_B$) and classical radii (sum) of electron and proton ($\alpha^2 a_B$), where $\alpha$ is the fine structure constant.

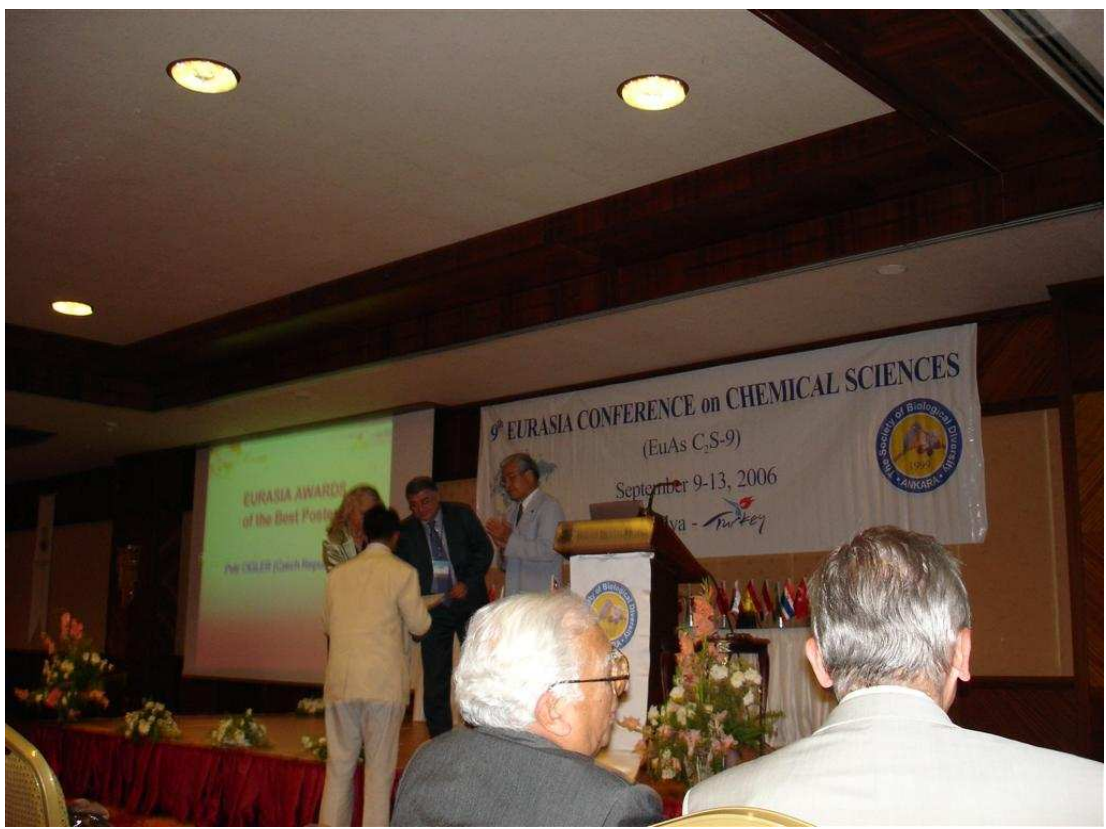

**Photo**: Professor Hitoshi Ohtaki (first from the right on the stage) at the 9th Eurasia Conference on Chemical Sciences, Antalya, Turkey, September 2006, who passed away on 5th November 2006.

## 1. Introduction

It was pointed out [1,2] about forty years ago, with the then available data, that the ratio of the van der Wals radii ($R_{vdW}$) of atoms to their de Broglie wavelengths ($\lambda_{dB}$) is nearly a constant for each group of non-metallic elements. The de Broglie wavelength ($\lambda_{dB} = 2\pi a_B$) is related to the ground state Bohr radius ($a_B$) and $a_B$ in turn to the ground state energy (or the ionization potential). Therefore, using the existing data [1, 3, 4], it is shown here that $R_{vdW}$ is indeed directly proportional to $a_B$ not only for the non-metallic elements, but also for all the A-group elements from 8A to 1A. The straight lines have different slopes and have non-zero intercepts (unlike in [1, 2]). In [4], which has data for elements of the groups 7A – 1A, two sets of data are given for the van der Waals radii, one from crystallographic data (which are comparable with Bondi's values [1, 3] for elements of groups 7A – 5A), and the other, denoted as equilibrium values are for the isolated atoms. The latter are larger since the atoms have more free space than in the crystalline lattices. For elements of other groups, the graphs are not linear and are not considered here.

The data given in [3] for the valence shell radii ($R_{v-s}$) and covalent radii $d(A)_{cov}$, show that they are also linearly dependent on $a_B$. Since the covalent radius $d(A)_{cov}$ is defined [6] as half the inter-atomic bond distance $d(AA)_{cov}$ between atoms of the same kind, the letter d is used here for the radius. The slopes and intercepts reproduce all the radii to about (+/-) 0.07 Å, which is within the scatter in the reported values.

Thus, presented below in sections 2 – 5 are new findings that there are linear correlations between $R_{vdW}$, $R_{v-s}$, $d(A)_{cov}$ and $a_B$. This shows that each of the radii is actually a sum of two lengths: one, a multiple of $a_B$, and the other, a constant for each group of elements from 8A – 1A.

## 2. Dependence of van der Waals radii ($R_{vdW}$) on the Bohr radii, $a_B$

The values of $R_{vdW}$ for various elements from [3, 4] are given here in columns 3, 5 and 7 in Table 1. The data in columns 5 and 7 are for crystals and equilibrium values for isolated atoms, respectively, from [4].

The first ionization potentials ($I_1$) from [3] were used for calculating $a_B$ from the relation,

$$a_B = e/2\kappa I_1 \qquad (1)$$

where e is the charge and $\kappa$ is the electrical permittivity of vacuum; see data in column 2, Table 1. (It is interesting to note from [5] that for many groups of elements in the Periodic Table, the aqueous redox potentials depend on $1/I_1 = 2\kappa a_B/e$). The linear dependence of $R_{vdW}$ on $a_B$ for elements of various groups are shown in Fig. 1 for the data from [3] and in Fig. 2 for data from [4] (in order to avoid too much overlapping and crowding, the lines corresponding to equilibrium values are not shown). Fig. 2 includes elements from groups 1A – 7A. For elements of group 3A, the points are a bit scattered. The straight lines are expressed here by the equations of the type $Y = Y^* + c$, where $Y^* = ma_B$, m and c are the slope and intercept respectively. Thus,

$$R_{vdW} = R^*_{vdW} + c_{vdW} \qquad (2a)$$

$$R^*_{vdW} = m_{vdW} a_B < R_{vdW} \qquad (2b)$$

The values of $m_{vdW}$ (> 1) and $c_{vdW}$ (< 1) are given in Table 1. The values of $R^*_{vdW}$ are tabulated in columns 4, 6 and 8 in Table 1. For the crystallographic and equilibrium values of $R_{vdW}$ from [4], the values of $R^*_{vdW}$ are mostly close to each other, but the intercepts c differ.

Thus, $R_{vdW}$ actually consists of the two terms, $R^*_{vdW}$, which depends on $a_B$, and $c_{vdW}$, a constant for each group of elements.

## 3. Dependence of valence shell radii, $R_{v-s}$ on the Bohr radii, $a_B$

The valence shell radii $R_{v-s}$ of atoms defined [3, 7] as 'the radius at which the magnitude of wave function is greatest for the valence orbitals', tabulated in [3] are presented here in column 9, Table 1. These radii are also linearly dependent on $a_B$, as can be seen from Fig. 3. They can be represented by equations similar to Eqs. (2),

$$R_{v-s} = R^*_{v-s*} + c_{v-s} \qquad (3a)$$

$$R^*_{v-s*} = m_{v-s} a_B > R_{v-s} \qquad (3b)$$

where the slopes, $m_{v-s} > 1$ and the intercepts, $c_{v-s}$ have negative values. This shows that the Bohr radius dependent part of the valence-shell radius, $R^*_{v-s} > R_{v-s}$ and is nearly $2a_B$ (= $\lambda_{dB}/\pi$) (the major axis of the Sommerfeld ellipse) for atoms of many groups. Note that the point for H belongs to the straight line for the alkali metals of group 1A.

## 4. Dependence of the covalent radii, $d(A)_{cov}$ on the Bohr radii, $a_B$

The data for the covalent radii $d(A)_{cov}$ from [3] are given in column 11, Table 1. Fig. 4 shows that $d(A)_{cov}$ is also linear with $a_B$ as in Figs. 1 - 3. The straight lines can be represented by,

$$d(A)_{cov} = d(A)^*_{cov} + c_A \qquad (4a)$$

$$d(A)^*_{cov} = m_A a_B > d(A)_{cov} \text{ (except for group 8A)} \qquad (4b)$$

The values of the slopes $m_A$ and intercepts $c_A$ and of $d(A)^*_{cov}$ are in Table 1. For elements of most of the groups (7A, 4A, 2A and 1A), $m_A$ is close to 2 and $d(A)^*_{cov} \sim 2a_B$ ($\sim R^*_{v-s}$), whereas for those of the inert gases, $m_A \sim 2.4$ and for elements of groups 6A and 5A, $m_A \sim 3$. Except for group 8A atoms, the intercepts $c_A$ are negative and the Bohr radius dependent part, $d(A)^*_{cov} > d(A)_{cov}$. Note again that, as in Fig. 3, the point for H falls on the same straight line as for the group 1A alkali metals.

Fig. 5 (drawn to scale) shows the relative comparisons of the Bohr radii ($a_B$) and the pairs of radii, $R^*_{vdW}$ & $R_{vdW}$, $R^*_{v-s}$ & $R_{v-s}$ and $d(A)^*_{cov}$ & $d(A)_{cov}$ for elements of groups 8A – 1A.

## 5. Dependences of $R_{vdW}$ and $R_{v-s}$ on $d(A)_{cov}$

Since $R_{vdW}$, $R_{v-s}$ and $d(A)_{cov}$ are all linearly dependent on $a_B$, both $R_{vdW}$ and $R_{v-s}$ are directly proportional to $d(A)_{cov}$. It was suggested in [6] that $R_{vdW}$ differs from $d(A)_{cov}$ by a constantt ~ 0.8 Å. Here, it is found that this difference depends $a_B$ as shown by,

$$R_{vdW} - d(A) = (m_{vdW} - m_A)a_B + (c_{vdW} - c_A) \qquad (5)$$

The values (see Table 1) show that the suggestion in [6] is a near approximation.

In conclusion, it can be seen that all the above radii (R) can be expressed in terms of the Bohr radii by a single linear equation,

$$R = ma_B + c \qquad (6)$$

Thus, the Bohr radius can be considered as a unit of length for the above radii as much as for the smaller Compton wavelength ($\alpha a_B$) and the sum of the classical radii of the proton and electron ($\alpha^2 a_B$), where $\alpha$ is the fine structure constant.

**Acknowledgment:** This research was supported by grants AVOZ50040507 of the Academy of Sciences of the Czech Republic and LC06035 of the Ministry of Education, Youth and Sports of the Czech Republic. The auhtor is gratfeul, to the Organizers of the 10[th] Eurasia Conference on Chemical Sciences, Manila, Philippines, January 2008, for inviting me with full financial support for presenting this work as an Ohtaki Memorial Lecture.

**Table 1.** Bohr radii, $a_B$, van der Waals radii, $R_{vdW}$, [3,4], valence shell radii, $R_{v-s}$, [3] and covalent radii, $d(A)_{cov}$, [3], ([6]: gp.3A) of atoms. (m,c: slope, intercept). All radii are in Å.

| Atom(Gp.) | $a_B$ | $R_{vdW}$[3] | $R^*_{vdW}$ | $R_{vdW}$[4] cryst. | $R^*_{vdW}$ cryst. | $R_{vdW}$[4] equil. | $R^*_{vdW}$ equil. | $R_{v-s}$ | $R^*_{v-s}$ | $d(A)_{cov}$ | $d(A)^*_{cov}$ |
|---|---|---|---|---|---|---|---|---|---|---|---|
| **8A: (m,c)** | | (2.56, 0.68) | | **cryst.** | cryst. | **equil.** | equil. | (1.93, -0.26) | | (2.37, 0.79) | |
| **He** | 0.29 | 1.40 | 0.75 | | | | | 0.30 | 0.56 | 1.50 | 0.69 |
| Ne | 0.33 | 1.54 | 0.85 | | | | | 0.36 | 0.64 | 1.57 | 0.79 |
| Ar | 0.46 | 1.88 | 1.17 | | | | | 0.63 | 0.88 | 1.86 | 1.08 |
| Kr | 0.51 | 2.02 | 1.32 | | | | | 0.73 | 0.99 | 2.02 | 1.22 |
| Xe | 0.59 | 2.16 | 1.52 | | | | | 0.91 | 1.14 | 2.20 | 1.40 |
| Rn | 0.67 | | | | | | | 1.00 | 1.29 | | |
| **7A: (m,c)** | | (1.86, 0.71) | | (2.15, 0.61) | | (2.08, 0.82) | | (1.96, -0.41) | | (2.16, -0.18) | |
| **F** | 0.41 | 1.47 | 0.77 | 1.50 | 0.89 | 1.65 | | 0.41 | 0.81 | 0.74 | 0.89 |
| Cl | 0.56 | 1.75 | 1.03 | 1.80 | 1.19 | 2.05 | | 0.68 | 1.09 | 0.99 | 1.20 |
| Br | 0.61 | 1.85 | 1.14 | 1.90 | 1.31 | 2.10 | | 0.77 | 1.19 | 1.15 | 1.32 |
| I | 0.69 | 1.98 | 1.28 | 2.10 | 1.48 | 2.22 | | 0.95 | 1.35 | 1.33 | 1.49 |
| **6A: (m,c)** | | (1.95, 0.48) | | (1.93, 0.50) | | (2.36, 0.45) | | (1.95, -0.58) | | (2.95, -0.98) | |
| **O** | 0.53 | 1.52 | 1.03 | 1.55 | 1.02 | 1.71 | 1.25 | 0.46 | 1.03 | 0.60 | 1.56 |
| S | 0.69 | 1.80 | 1.35 | 1.80 | 1.34 | 2.06 | 1.64 | 0.75 | 1.35 | 1.03 | 2.05 |
| Se | 0.74 | 1.90 | 1.44 | 1.90 | 1.42 | 2.18 | 1.74 | 0.82 | 1.44 | 1.16 | 2.18 |
| Te | 0.80 | 2.06 | 1.55 | 2.10 | 1.54 | 2.36 | 1.89 | 1.00 | 1.56 | 1.43 | 2.36 |
| Po | 0.86 | | | | | | | 1.09 | 1.67 | 1.67 | |
| **5A: (m,c)** | | (1.27, 0.92) | | (1.80, 0.71) | | (1.85, 0.87) | | (1.54, -0.23) | | (2.72, -0.78) | |
| **N** | 0.50 | 1.55 | 0.63 | 1.60 | 0.89 | 1.79 | 0.92 | 0.54 | 0.76 | 0.55 | 1.35 |
| P | 0.69 | 1.80 | 0.87 | 1.95 | 1.24 | 2.14 | 1.27 | 0.84 | 1.06 | 1.11 | 1.86 |
| As | 0.73 | 1.85 | 0.93 | 2.05 | 1.32 | 2.25 | 1.36 | 0.89 | 1.13 | 1.25 | 1.99 |
| Sb | 0.83 | | | 2.20 | 1.50 | 2.41 | 1.54 | 1.07 | 1.28 | 1.45 | 2.26 |
| **4A: (m,c)** | | | | (1.58, 0.69) | | (1.41,1.05) | | (1.48, -0.35) | | (1.98, -0.56) | |
| **C** | 0.64 | 1.70 | | 1.70 | 1.01 | 1.96 | 0.90 | 0.61 | 0.95 | 0.71 | 1.27 |
| Si | 0.88 | 2.10 | | 2.10 | 1.40 | 2.26 | 1.25 | 0.95 | 1.31 | 1.18 | 1.75 |
| Ge | 0.91 | | | 2.10 | 1.44 | 2.32 | 1.29 | 0.96 | 1.35 | 1.23 | 1.81 |
| Sn | 0.98 | | | 2.25 | 1.55 | 2.46 | 1.38 | 1.14 | 1.45 | 1.41 | 1.94 |
| **3A: (m,c)** | | | | (1.02, 0.92) | | (1.22, 1.00) | | (1.06, -0.10) | | (1.93, -0.76) | |
| B | 0.87 | | | 1.8 | 0.89 | 2.05 | 1.06 | 0.81 | 0.92 | 0.90 | 1.67 |
| Al | 1.20 | | | 2.1 | 1.23 | 2.4 | 1.47 | 1.11 | 1.28 | 1.43 | 2.32 |
| Ga | 1.20 | | | 2.1 | 1.22 | 2.41 | 1.46 | 1.06 | 1.27 | 1.40 | 2.32 |
| In | 1.24 | | | 2.2 | 1.27 | 2.53 | 1.52 | 1.23 | 1.32 | 1.69 | 2.40 |
| Tl | 1.18 | | | 2.2 | 1.20 | 2.53 | 1.44 | 1.30 | 1.25 | 1.73 | 2.27 |
| **2A: (m,c)** | | | | (1.25, 0.96) | | (1.40, 1.13) | | (2.06, -0.55) | | (2.11, -0.46) | |
| **Be** | 0.77 | | | 1.90 | 0.97 | 2.23 | 1.08 | 1.09 | 1.59 | 1.14 | 1.63 |
| Mg | 0.94 | | | 2.20 | 1.18 | 2.42 | 1.32 | 1.37 | 1.94 | 1.60 | 1.99 |
| Ca | 1.18 | | | 2.40 | 1.47 | 2.78 | 1.65 | 1.84 | 2.43 | 1.98 | 2.49 |
| Sr | 1.26 | | | 2.55 | 1.58 | 2.94 | 1.77 | 2.05 | 2.61 | 2.15 | 2.67 |
| Ba | 1.38 | | | 2.70 | 1.73 | 3.05 | 1.93 | 2.35 | 2.85 | 2.51 | 2.92 |
| **1A: (m,c)** | | | | (1.43, 0.39) | | (1.25, 0.99) | | (2.24, -1.36) | | (2.04, -0.71) | |
| H | 0.53 | 1.20 | | 1.17 | 0.76 | | | 0.53 | | 0.37 | 1.08 |
| **Li** | 1.34 | 1.82 | | 2.20 | 1.91 | 2.63 | 1.67 | 1.64 | 2.99 | 2.00 | 2.72 |
| Na | 1.40 | 2.27 | | 2.40 | 2.00 | 2.77 | 1.75 | 1.79 | 3.14 | 2.15 | 2.86 |
| K | 1.66 | 2.75 | | 2.80 | 2.37 | 3.02 | 2.07 | 2.30 | 3.72 | 2.66 | 3.38 |
| Rb | 1.72 | | | 2.90 | 2.46 | 3.15 | 2.15 | 2.49 | 3.86 | 2.79 | 3.51 |
| Cs | 1.85 | | | 3.00 | 2.64 | 3.30 | 2.31 | 2.82 | 4.15 | 3.07 | 3.77 |

**Legends for 5 Figures.**

**Fig. 1.** Linear dependence of van der Waals radii, $R_{vdW}$ from [3] on the ground state Bohr radii, $a_B$, for elements of groups 8A - 5A; see Eq. (2a).

**Fig. 2.** Linear dependence of van der Waals radii, $R_{vdW}$ from [4] on the ground state Bohr radii, $a_B$, for elements of groups 7A - 1A; see Eq. (2a). Note that the point for H is on the same line as for Group 1A elements.

**Fig. 3.** Linear dependence of valence shell radii, $R_{v-s}$ [3] on the ground state Bohr radii $a_B$, for elements of groups 8A – 1A; see Eq. (3a).

**Fig. 4.** Linear dependence of covalent radii $d(A)_{cov}$ [3] on the ground state Bohr radii for elements of groups 8A – 1A; see Eq. (4a). For Li [3], $d(A)_{cov}$ is half the diagonal of a square with the side equal to the length from the body center to the corner of the cube. Note that the point for H is on the same line as for Group 1A elements.

**Fig. 5.** Comparison of the radii $a_B$, $R^*_{vdW}$ & $R_{vdW}$, $R^*_{v-s}$ & $R_{v-s}$, $d(A)^*_{cov}$ and $d(A)_{cov}$ (as radii of circles) for elements of groups 8A – 1A. Data for $R_{vdW}$ from [4] (from crystallographic data) have been used for groups 6A - 1A and all other data are from [3]; see Table 1. Note the gradations in the radii: $a_B < R_{v-s} < R_{vdW} < d(A)_{cov}$ (group 8A); $a_B < R_{v-s} < R_{vdW} > d(A)_{cov}$ (groups 7A - 4A and 2A); $a_B \sim R_{v-s} < R_{vdW} > d(A)_{cov}$ (group 3A); $a_B < R_{v-s} < R_{vdW} \sim d(A)_{cov}$ (group 1A).

Fig. 1 (R.H.)

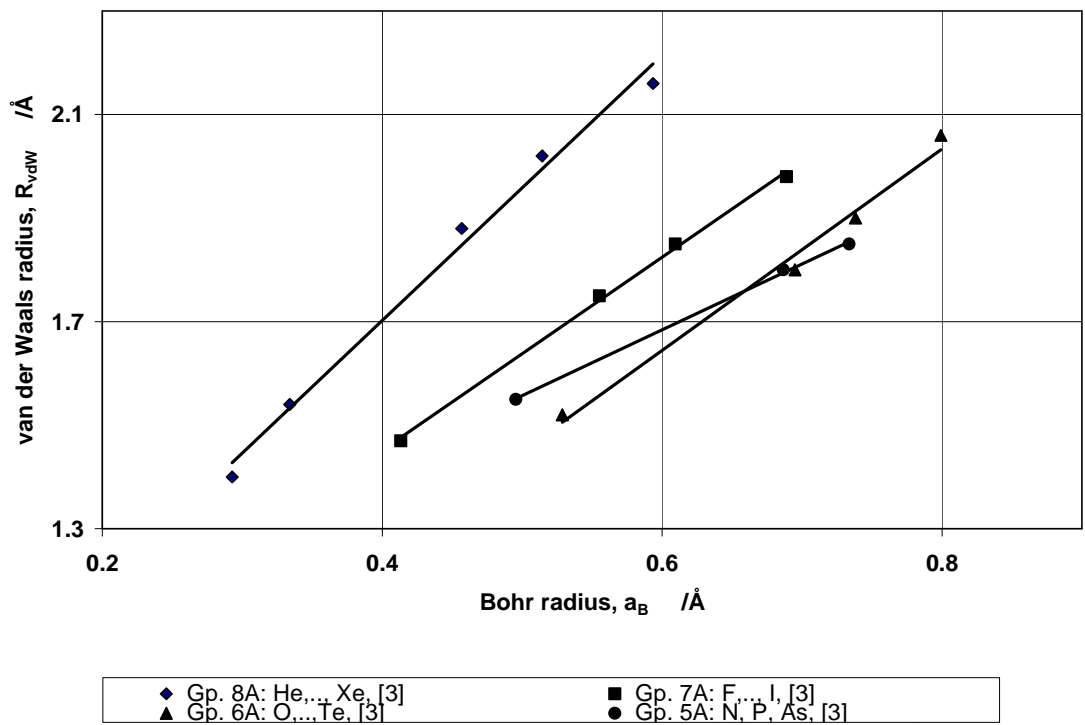

| ♦ Gp. 8A: He,..., Xe, [3] | ■ Gp. 7A: F,..., I, [3] |
| ▲ Gp. 6A: O,...,Te, [3] | ● Gp. 5A: N, P, As, [3] |

Fig. 2 (R.H.)

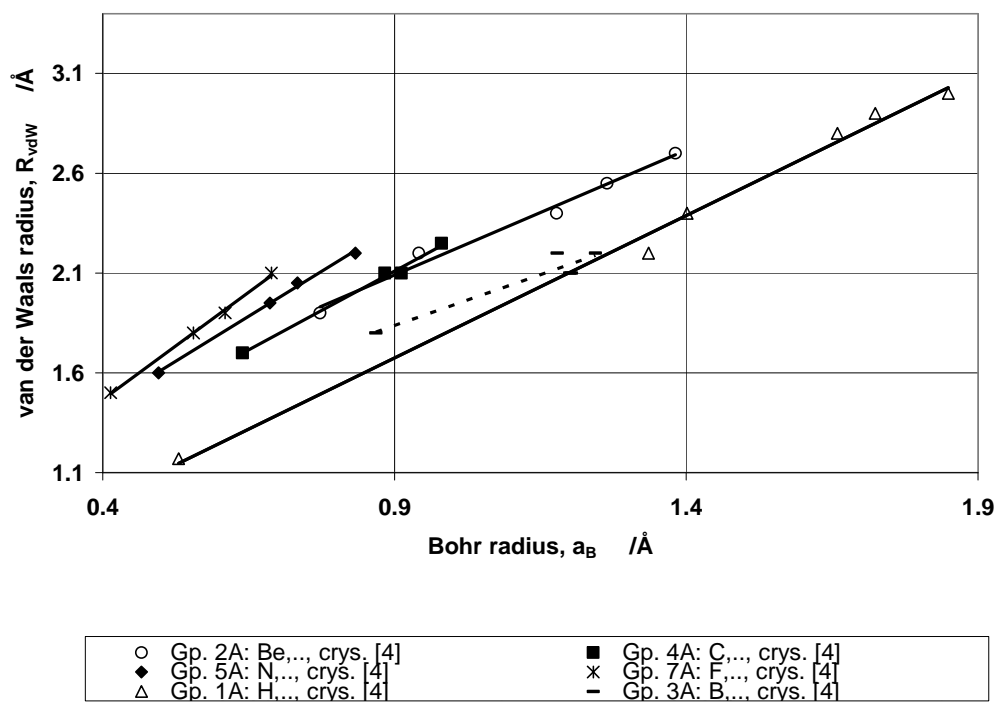

| ○ Gp. 2A: Be,..., crys. [4] | ■ Gp. 4A: C,..., crys. [4] |
| ♦ Gp. 5A: N,..., crys. [4] | ✶ Gp. 7A: F,..., crys. [4] |
| △ Gp. 1A: H,..., crys. [4] | – Gp. 3A: B,..., crys. [4] |

Fig. 3 (R.H.)

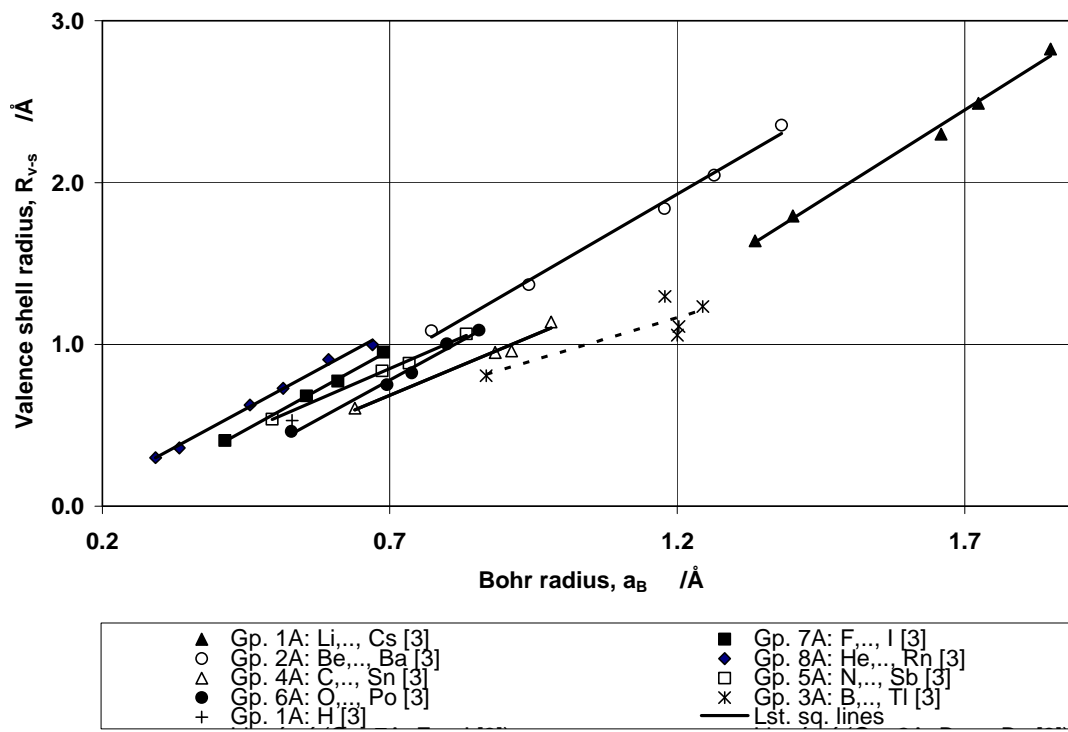

Fig. 4 (R.H.)

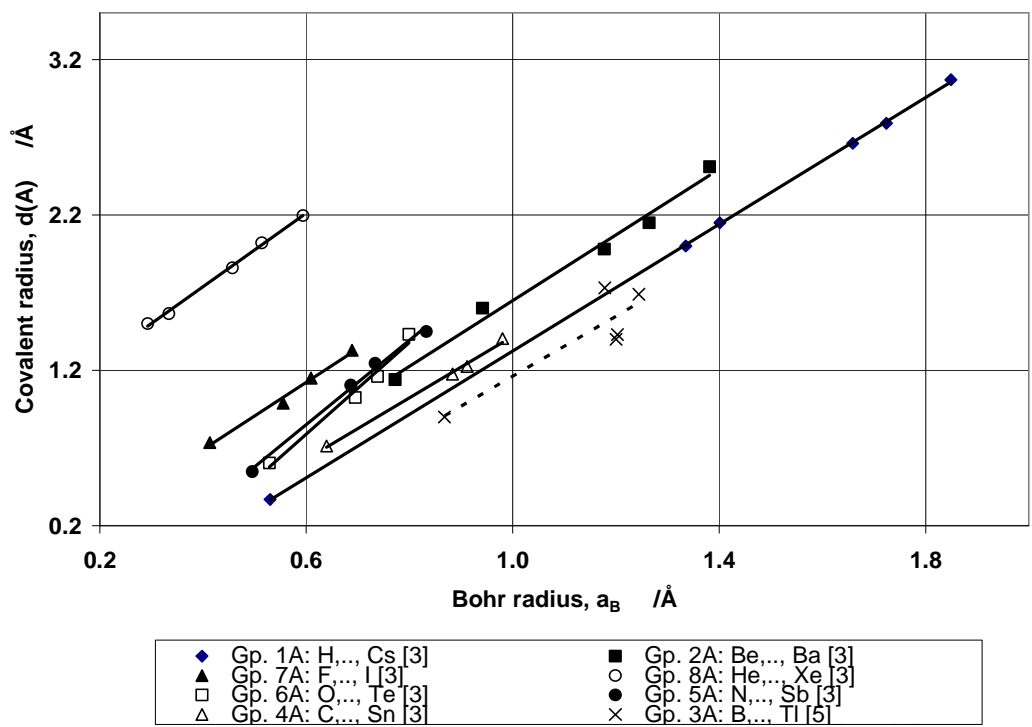

Fig. 5 (R.H.)

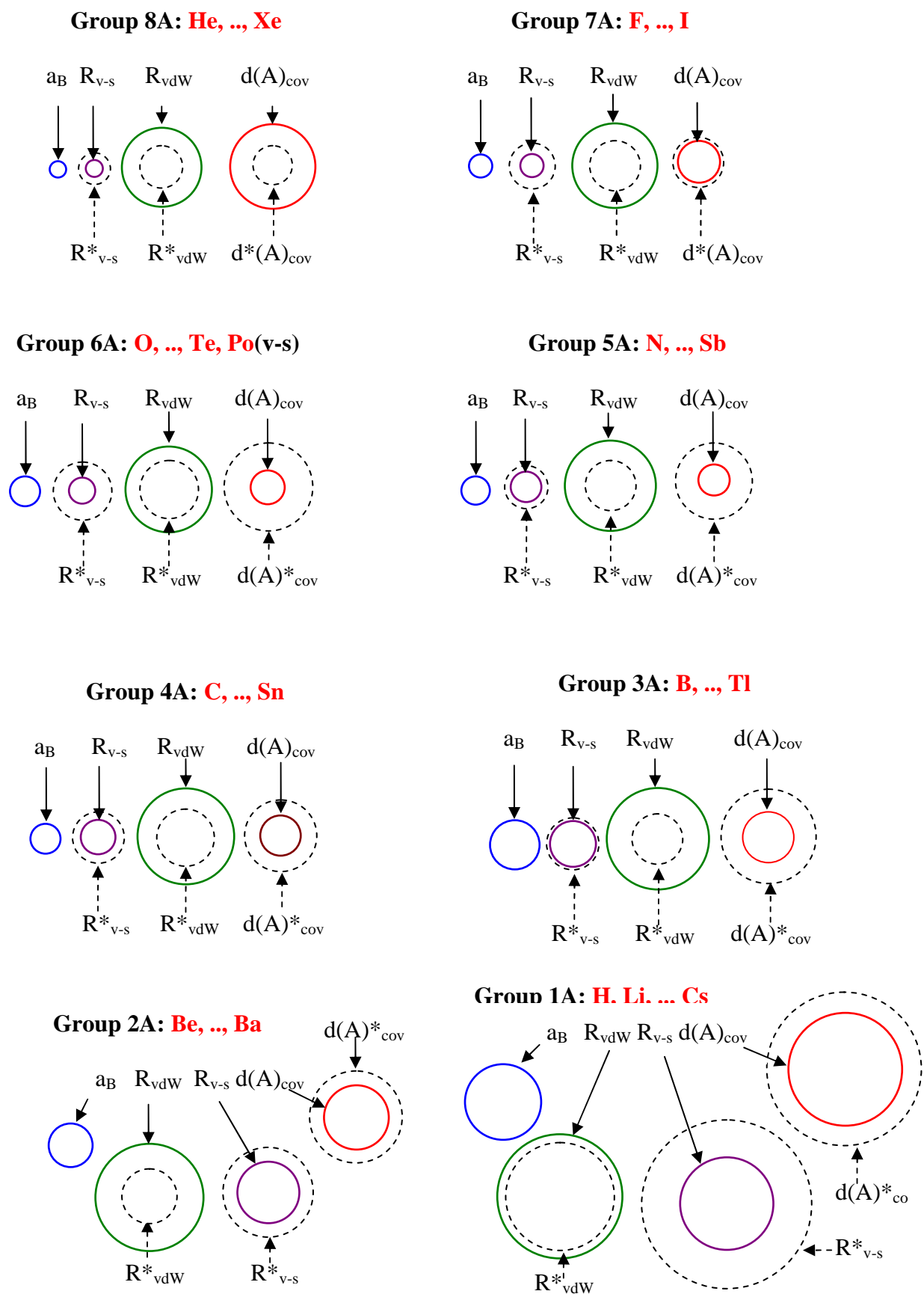